\documentclass[a4paper,copyright,creativecommons,noderivs]{eptcs}
\providecommand{\event}{Gandalf 2025} %

\usepackage{iftex}

\ifpdf
  \usepackage{underscore}         %
  \usepackage[T1]{fontenc}        %
\else
  \usepackage{breakurl}           %
\fi

\usepackage{amssymb,amsmath,amsthm}
\usepackage{tikz}
\usetikzlibrary{automata,calc}
\newtheorem{theorem}{Theorem}
\newtheorem{lemma}{Lemma}
\newtheorem{proposition}{Proposition}
\newtheorem{definition}{Definition}

\newtheorem{remark}{Remark}

\newcommand{\NN}{\mathbb{N}}
\newcommand{\hdcw}{HD-tCBW}

\title{How Concise are Chains of co-Büchi Automata?\thanks{Funded by Volkswagen Foundation within its \emph{Momentum} framework under project no.~9C283}}
\author{Rüdiger Ehlers
\institute{Clausthal University of Technology}
\email{ruediger.ehlers@tu-clausthal.de}
}

\newcommand{\titlerunning}{How Concise are Chains of co-Büchi Automata?}
\newcommand{\authorrunning}{R.~Ehlers}

\hypersetup{
  bookmarksnumbered,
  pdftitle    = {\titlerunning},
  pdfauthor   = {Rüdiger Ehlers},
  pdfsubject  = {EPTCS},               %
}

\begin{document}
\maketitle

\begin{abstract}
Chains of co-Büchi automata (COCOA) have recently been introduced as a new canonical model for representing arbitrary $\omega$-regular languages. They can be minimized in polynomial time and are hence an attractive language representation for applications in which normally, deterministic $\omega$-automata are used.
While it is known how to build COCOA from deterministic parity automata, little is currently known about their relationship to automaton models introduced earlier than COCOA.

In this paper, we 
analyze the conciseness of chains of co-Büchi automata. We 
show that even in the case that all automata in the chain are deterministic, chains of co-Büchi automata can be exponentially more concise than deterministic parity automata. %
We then answer the question if this conciseness is retained when performing Boolean operations (such as disjunction and conjunction) over COCOA by showing that there exist families of languages for which these operations lead to an exponential growth of the sizes of the automata. The families have the property that when representing them using deterministic parity automata, taking the disjunction or conjunction of them only requires a polynomial blow-up, which shows that Boolean operations over COCOA do not retain their conciseness in general. 
\end{abstract}

\section{Introduction}

Automata over infinite words are a classical model for representing the specification of a reactive system.
They augment temporal logics such as linear temporal logic (LTL, \cite{DBLP:conf/focs/Pnueli77}) and linear dynamic logic (LDL, \cite{DBLP:conf/ijcai/GiacomoV13,DBLP:journals/iandc/FaymonvilleZ17}) by providing an intermediate representation for a specification that is structured in a way so that it can be used directly in verification and synthesis algorithms.
While for classical model checking of finite-state systems, non-deterministic automata with a Büchi acceptance condition suffice, for some applications, such as reactive synthesis and probabilistic model checking, richer automata  types are employed. 
In this context, deterministic automata with parity acceptance are particularly interesting as when a specification is given as such, the reactive synthesis problem over the specification can be reduced to solving a parity game based on the state space structure of the automaton \cite{DBLP:reference/mc/BloemCJ18}.

Unfortunately, deterministic parity automata can become quite large in practice, which complicates employing them in reactive synthesis. For instance, when translating an LTL formula to a deterministic parity automaton, a doubly-exponential blow-up cannot be avoided in the worst case \cite{DBLP:conf/mochart/KupfermanR10}. However, even for languages that do not require such huge automata, current translation procedures for obtaining deterministic parity automata can compute unnecessarily large automata, caused by them only applying heuristics for size minimization. Given that deterministic parity automaton minimization is NP-hard \cite{DBLP:conf/fsttcs/Schewe10,DBLP:journals/corr/abs-2504-20553}, this is not surprising.

To counter this problem, chains of co-Büchi automata (COCOA) have recently been proposed as a new model for $\omega$-regular languages \cite{DBLP:conf/fsttcs/EhlersS22}.
In a COCOA, the language to be represented is split into a falling chain of co-Büchi languages, where each of the co-Büchi languages is represented as a \emph{history-deterministic} co-Büchi automaton with \emph{transition-based acceptance} (\hdcw). In this context, tran\-si\-tion-based acceptance refers to the transitions being accepting or rejecting rather than the states. This particular type of co-Büchi automata is minimizable in polynomial time \cite{DBLP:journals/lmcs/RadiK22}, so that each automaton in the chain can be minimized separately. To employ these automata in a canonical and polynomial-time minimizable model for arbitrary $\omega$-regular languages, a canonical split of an $\omega$-regular language to co-Büchi automata was defined \cite{DBLP:conf/fsttcs/EhlersS22}. COCOA can be thought of as assigning a \emph{color} to each word, just as deterministic parity automata do. A word then has a color of $i$ (for some $i \in \NN$) if the $i$th automaton in the chain accepts the word, but no automaton later in the chain accepts the word. The core contribution of the COCOA definition is a concretization of which color should be assigned to each word, and this concretization is not based on some automaton representation of the language, but only on the language itself, called the \emph{natural color} of the respective word.

COCOA have already found first use in reactive synthesis \cite{DBLP:conf/tacas/EhlersK24}, based on a procedure for translating deterministic parity automata to COCOA \cite{DBLP:conf/fsttcs/EhlersS22}. Given that polynomial-time minimization is an attractive property for future applications as well, it makes sense to have a closer look at the properties of COCOA and their relationship to earlier automata types, in particular in relation to deterministic parity automata, which they have the potential of replacing in some applications. For instance, to understand when they are a suitable specification model and to inform the future development of procedures for manipulating COCOA, their \emph{conciseness} in relation to deterministic parity automata needs to be understood.

In this paper, we provide a study of the conciseness of COCOA  with a particular focus on deterministic parity automata as comparison basis. Apart from summarizing how some existing results on deterministic co-Büchi automata transfer to the COCOA case, we provide two new COCOA-specific technical results:
\begin{enumerate}
\item 
We show that COCOA can be exponentially more concise than deterministic parity automata (DPW) even when the co-Büchi languages in the COCOA are representable as small deterministic co-Büchi automata and when the overall language only has one residual language.
{While it was previously known that COCOA can be exponentially more concise than deterministic parity automata, this was due to the automata in the COCOA being history-deterministic, and \hdcw{} are known to be exponentially more concise than deterministic automata (for some languages). The new result in this paper shows that COCOA can be exponentially more concise than DPW even when not making use of history-determinism for the chain elements.}

\item We show that exponential conciseness of COCOA over deterministic parity automata can be lost when performing Boolean operations (such as conjunction or disjunction) on COCOA. In particular, such Boolean operations can require an exponential growth in the number of states even in cases in which for deterministic parity automata, such a growth is not necessary.
\end{enumerate}
These results shed light on the fundamental properties of COCOA. In the first case, the example family of languages defined for the result shows that even with small automata in a chain, complex liveness languages can be composed. The second example shows that the property of a COCOA to have a number of residual languages that is exponential in their size can be lost in the case of Boolean operations. It hence demonstrates that future procedures for performing Boolean operations on COCOA will need to have an exponential lower bound on the sizes of the resulting COCOA.

Both main technical results involve carefully defining families of COCOA that exemplify the respective lower bounds. For each family, we have to show that the COCOA given indeed recognize every word with their respective natural color, which requires substantial care. 

After stating some preliminaries, we give a summary of the ideas behind COCOA in Section~\ref{sec:COCOA}. Section~\ref{sec:conciseness} summarizes the implications of known results on the conciseness of COCOA and provides the first new technical result. The section afterwards contains the lower bound on the blow-up incurred by Boolean operations on COCOA. The paper closes with a discussion of the obtained results in Section~\ref{sec:conclusion}.

\section{Preliminaries}

\textbf{Languages:} For a finite set $\Sigma$ as \emph{alphabet}, let $\Sigma^*$ denote the set of finite words over $\Sigma$, and $\Sigma^\omega$ be the set of infinite words over $\Sigma$. A subset $L \subseteq \Sigma^\omega$ is also called a \emph{language}. Given a language $L$ and some finite word $w \in \Sigma^*$, we say that $L|_w = \{w' \in \Sigma^\omega \mid ww' \in L\}$ is the \emph{residual language} of $L$ over $w$. Given a word $w = w_0 w_1 \ldots \in \Sigma^\omega$ and some language $L$, we say that an infinite word $w' = w_0 w_1 \ldots w_i \tilde w w_{i+1} w_{i+2} \ldots$ results from a \emph{residual language invariant injection} of $\tilde w$ at position $i \in \NN$ if $L|_{w_0 \ldots w_i} = L|_{w_0 \ldots w_i \tilde w}$.

\textbf{Automata:} Some languages, in particular the \emph{$\omega$-regular languages}, can be represented by \emph{parity automata}. We only consider automata with \emph{transition-based acceptance} in this paper. These are tuples of the form $\mathcal{A} = (Q,\Sigma,\delta,q_0)$ in which $Q$ is a finite set of states, $\Sigma$ is the alphabet, $q_0 \in Q$ is the initial state of the automaton, and $\delta \subseteq Q \times \Sigma \times Q \times \NN$ is its \emph{transition relation}.

Given a word $w = w_0 w_1 \ldots \in \Sigma^\omega$, we say that $w$ induces an infinite run $\pi = \pi_0 \pi_1 \ldots \in Q^\omega$ together with a sequence of \emph{colors} $\rho = \rho_0 \rho_1 \ldots \in \NN^\omega$ if we have $\pi_0 = q_0$ and for all $i \in \NN$, we have $(\pi_i,w_i,\pi_{i+1},\rho_i) \in \delta$. In this paper, we only consider automata that are \emph{input-complete}, i.e., for which for each state/letter combination $(q,x)$, there exists at least one pair $(q',c)$ with $(q,x,q',c) \in \delta$. We furthermore only consider automata for which the color $c$ does not depend on the transition taken, so that for each $(q,x)$, there is only one value $c$ with $(q,x,q',c) \in \delta$ for some $q'$.

A run is accepting if for the corresponding color sequence $\rho$ (which is unique), we have that the lowest color occurring infinitely often in it is even. This color is also called the \emph{dominating color} of the run. A word is accepted by $\mathcal{A}$ if there exists an accepting run for it. The language of $\mathcal{A}$, written $\mathcal{L}(\mathcal{A})$, is the set of words with accepting runs.
An automaton is said to be deterministic if for every $(q,x) \in Q \times \Sigma$, there exists exactly one combination $(q',c) \in Q \times \NN$ with $(q,x,q',c) \in \delta$. In such a case, we also refer to the dominating color of the unique run as the color with which the automaton \emph{recognizes} the word. We say that $\mathcal{A}$ is a \emph{co-Büchi automaton} if the only colors occurring along transitions in $\mathcal{A}$ are 1 and 2. A \emph{co-Büchi language} is a language of some co-Büchi automaton. 

We say that an automaton $\mathcal{A}$ represents the \emph{disjunction} of some automata $\mathcal{A}_1$ and $\mathcal{A}_2$ if $\mathcal{L}(\mathcal{A}) = \mathcal{L}(\mathcal{A}_1) \cup \mathcal{L}(\mathcal{A}_2)$. It represents the \emph{conjunction} of  $\mathcal{A}_1$ and $\mathcal{A}_2$ if $\mathcal{L}(\mathcal{A}) = \mathcal{L}(\mathcal{A}_1) \cap \mathcal{L}(\mathcal{A}_2)$. {The \emph{size} of an automaton is defined to be the number of its states.}

\looseness-1 \textbf{History-deterministic automata:} Parity automata, as defined above, are not necessarily deterministic. We consider \emph{history-deterministic} co-Büchi automata (\hdcw{}) in particular. For them, there exists some \emph{advice} function $f : \Sigma^* \rightarrow Q$ such that for each word, if and only if $w = w_0 w_1 \ldots \in \mathcal{L}(\mathcal{A})$, the sequence $q_0 f(w_0) f(w_1 w_2 \ldots) \ldots$ is a valid accepting run of the automaton. Abu Radi and Kupferman~\cite{DBLP:journals/lmcs/RadiK22} showed how to minimize such automata, and in minimized automata, for every state/letter combination, all transitions have the same color (so that the assumption from above is justified). History-deterministic co-Büchi automata are also called \emph{good-for-games} co-Büchi automata in the literature. The sets of languages representable by \hdcw{} and deterministic co-Büchi automata are the same. Deterministic parity automata (DPW) are however strictly more expressive.
We also sometimes represent automata in a graphical notation, where states are circles, transitions are arrows between circles, and the initial state is marked by an arrow from a dot. In co-Büchi automata, dashed arrows represent \emph{rejecting transitions} (with color $1$), while the solid arrows represent \emph{accepting transitions} (with color $2$). For parity automata, the edges are labeled by the color numbers in addition to their alphabet letters.

\looseness-1 \textbf{SCCs:} Given an automaton $\mathcal{A} = (Q,\Sigma,\delta,q_0)$, we say that some tuple $(Q',\delta')$ with $Q' \subseteq Q$ and $\delta' \subseteq \delta$ is a \emph{strongly connected component} (SCC) of $\mathcal{A}$ if for each $q, q' \in Q'$, there exists a sequence of transitions within $\delta'$ for reaching $q'$ from $q$. Similarly, every transition within $\delta'$ is used in some such sequence.

{\textbf{Temporal logic and $\omega$-regular expressions:}
\emph{Linear temporal logic} (LTL, \cite{DBLP:conf/focs/Pnueli77}) is a formalism for expressing (some) languages over $\Sigma = 2^\mathsf{AP}$ for a set $\mathsf{AP}$. It is known that LTL can be translated to deterministic parity automata of size doubly-exponential in the size of the LTL property, and this blow-up bound is tight (see, e.g., \cite{DBLP:conf/tacas/EsparzaKRS17}).
We also use \emph{$\omega$-regular expressions} for stating some languages. These extend classical regular expressions by a symbol for infinite repetition, namely $^\omega$.
}

\section{A short introduction to chains of co-Büchi automata}
\label{sec:COCOA}

Chains of co-Büchi automata (COCOA, used as both the singular and plural form) provide a canonical representation for arbitrary $\omega$-regular languages.
Let a language $L$ over an alphabet $\Sigma$ be given. The starting point of a COCOA representation of $L$ is the decomposition of $\Sigma^\omega$ into a chain of languages $L_1 \supset L_2 \supset \ldots \supset L_n$.
A word $w$ is in the language represented by the chain, also denoted as $\mathcal{L}(L_1, \ldots, L_n)$ henceforth, if the \emph{highest} index $i$ such that $w \in L_i$ is even or $w \notin L_1$.
Each language $L_i$ (for some $1 \leq i \leq n$) represents the set of words whose \emph{natural color} (with respect to $L$) is at least $i$. The natural color of a word is the minimal color in which a word is \emph{at home}, which in turn is defined as follows:

\begin{definition}[\cite{DBLP:conf/fsttcs/EhlersS22}, Def.~1]
\label{def:naturalColorsOfWords}
Let $L$ be a language and $i \in \NN$. We say that a word $w$ is at home in a color of $i$ if there exists a sequence of injection points $J \subset \NN$ such that for all words $w'$ that result from injecting residual language invariant words at word positions in $J$, {we either have:
\begin{itemize}
\item $w'$ is at home in a color strictly smaller than $i$, or
\item both $w$ and $w'$ are in $L$ and $i$ is even, or both $w$ and $w'$ are not in $L$ and $i$ is odd.
\end{itemize}
Note that in the case of $i=0$, only the second case can} apply.
\end{definition}
The concept of the natural color of a word generalizes the idea of colors in a parity automaton in a way that is agnostic to the concrete choice of automaton for representing the language. The inductive definition above starts from color $0$, so that the languages at each level are uniquely defined.

\looseness-1 With this definition, not only is a chain $L_1, \ldots, L_n$ of languages uniquely defined for each $\omega$-regular language $L$, but we also have that for each $1 \leq i \leq n$, the language $L_i$ is a co-Büchi language \cite{DBLP:conf/fsttcs/EhlersS22}, i.e., it can be represented by a co-Büchi automaton. Hence, we can represent the chain of languages $L_1, \ldots, L_n$ by a chain of history-deterministic co-Büchi automata $A_1, \ldots, A_n$. Each of these automata can be minimized and made canonical in polynomial time \cite{DBLP:journals/lmcs/RadiK22}. 
Since the representation of $L$ as a chain of co-Büchi languages $L_1, \ldots, L_n$ is also canonical, we %
obtain a canonical representation of $L$. 

Details on the COCOA language representation can be found in the paper introducing COCOA \cite{DBLP:conf/fsttcs/EhlersS22} and in a video recording of a presentation of the paper's concepts with  additional examples
\cite{naturalColorsVideo}. 

\subsection{An example COCOA}

\begin{figure}
\centering\begin{tikzpicture}

\begin{scope}[xshift=-1cm]
\node[state,thick,fill=black!10!white] (a) at (0,0) {$q_0$};
\draw[thick,->,dashed] (a) to[loop above] node[above] {$a$} (a);
\draw[thick,->] (a) to[loop left] node[left] {$b,c$} (a);
\draw[fill=black] ($(a)+(-0.7,0.6)$) circle (0.05cm);
\draw[->,thick] ($(a)+(-0.7,0.6)$) -- (a);

\end{scope}

\node[state,thick,fill=black!10!white] (b) at (3,0) {$q_1$};
\node[state,thick,fill=black!10!white] (c) at (5,0) {$q_2$};
\draw[thick,->,dashed] (b) to[loop above] node[above] {$a$} (b);
\draw[thick,->,dashed] (c) to[loop above] node[above] {$a,b$} (c);
\draw[thick,->,dashed] (c) to[bend right=60] node[above] {$a,b$} (b);
\draw[thick,->] (c) to[bend left=0] node[below] {$c$} (b);
\draw[thick,->] (b) to[bend left=20] node[above] {$b$} (c);
\draw[thick,->] (b) to[loop left] node[left] {$c$} (b);
\draw[fill=black] ($(b)+(-0.7,0.6)$) circle (0.05cm);
\draw[->,thick] ($(b)+(-0.7,0.6)$) -- (b);

\begin{scope}[xshift=1cm]
\node[state,thick,fill=black!10!white] (d) at (8,0) {$q_3$};
\node[state,thick,fill=black!10!white] (e) at (10,0) {$q_4$};
\draw[thick,->,dashed] (d) to[loop above] node[above] {$b$} (d);
\draw[thick,->,dashed] (e) to[loop above] node[above] {$b,c$} (e);
\draw[thick,->] (d) to[loop left] node[left] {$c$} (d);
\draw[thick,->,dashed] (e) to[bend left=0] node[below] {$a$} (d);
\draw[thick,->,dashed] (d) to[bend left=20] node[above] {$a$} (e);
\end{scope}

\draw[fill=black] ($(d)+(-0.7,0.6)$) circle (0.05cm);
\draw[->,thick] ($(d)+(-0.7,0.6)$) -- (d);

\node at ($(a)+(0,-0.9)$) {$A_1$};
\node at ($0.5*(b)+0.5*(c)+(0,-0.9)$) {$A_2$};
\node at ($0.5*(d)+0.5*(e)+(0,-0.9)$) {$A_3$};

\end{tikzpicture}
\label{fig:exampleCOCOA}
\caption{An example COCOA}
\end{figure}
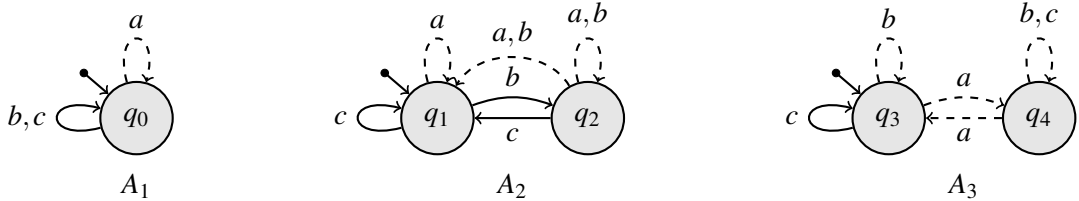

Figure~\ref{fig:exampleCOCOA} shows an example COCOA consisting of three automata, all over the alphabet $\Sigma = \{a,b,c\}$, together representing some language $L$. The chain's language contains the words with an infinite number of $a$ letters (with a natural color of $0$) as well as words that satisfy three conditions:
\begin{itemize}
\item the word ultimately only consists of $b$s and $c$s,
\item eventually, every $b$ is immediately followed by a $c$, and
\item if there is a finite even number of $a$ letters in the word, there are infinitely many $b$s.
\end{itemize}
Words satisfying these three conditions but only having finitely many $a$s have a natural color of $2$. Words not in $L$ have natural colors of $1$ or $3$. The COCOA hence recognizes words with four different natural colors, and it follows from the existing translation procedure from deterministic parity automata to COCOA \cite{DBLP:conf/fsttcs/EhlersS22} that every deterministic parity automaton for this language also needs at least four colors. 
The colors represent how often by injecting residual language invariant finite words an infinite number of times, words can alternate between being in $L$ or not.
In this example, the word $c^\omega$ has color $3$ and is hence not in $L$. By injecting $b$s such that the resulting word never has two $b$ letters in a row, the word becomes contained in $L$. By injecting $bb$ infinitely often, the word  leaves $L$ again. Finally, by injecting $a$ infinitely often, the final word is rejected by $A_1$ and hence in $L$. The overall language $L$ represented by the COCOA has two residual languages, but only $A_3$ tracks them and not $A_2$ or $A_1$. The relevance of the injection point set $J$ in Definition~\ref{def:naturalColorsOfWords} is not exemplified in the COCOA in Figure~\ref{fig:exampleCOCOA}, as for all COCOA discussed in the following sections, this set can be freely chosen and is hence not of relevance.

\subsection{Some additional definitions and notes in the context of COCOA}

For convenience, whenever we are dealing with a COCOA $A_1, \ldots, A_n$ in the following, we will assume that $A_0 = \Sigma^\omega$ and $A_{n+1} = \emptyset$, as this avoids dealing with special cases in some constructions while not affecting the definition of the COCOA's language.
To avoid cluttering the exposition in the following, the word \emph{injection} always refers to a residual language invariant word injection. When a word $w'$ is the result of a residual language invariant word injection into some word $w$, we say that $w'$ \emph{extends} $w$.
We define the sum of the automaton sizes in a COCOA to be the size of the COCOA.

The condition for a chain of co-Büchi automata  $A_1, \ldots, A_n$ to represent a language $L$ can be equivalently stated as requiring $A_1$ to reject the words with a natural color of $0$ (with respect to $L$) and that for each $1 \leq i \leq n$, the words accepted by $A_i$ but rejected by $A_{i+1}$ (if $i<n$) are the ones with a natural color of $i$ (with respect to $L$).
The definitions above also imply that the natural language of a word $w$ can only decrease by injecting letters into $w$ {(if the set $J$ of positions to inject at is chosen according to the requirements of Definition~\ref{def:naturalColorsOfWords})}.

\section{On the conciseness of COCOA}
\label{sec:conciseness}

In this section, we will relate the sizes of deterministic parity automata to the sizes of COCOA (for the same languages).
We summarize the implications of existing results on the conciseness of COCOA and augment them by new insights.

\paragraph{DPW conciseness over COCOA:} For starters, the translation by Ehlers and Schewe \cite{DBLP:conf/fsttcs/EhlersS22} for obtaining a COCOA from a deterministic parity automaton with $n$ states and $c$ colors yields COCOA with at most $c$ history-deterministic co-Büchi automata, each having at most $n$ many states. Even more, since the construction by Ehlers and Schewe minimizes the numbers of colors on-the-fly, this fact also holds for $c$ being the minimal number of colors that \emph{any} DPW for the language has. Hence, a COCOA can only be polynomially larger than a deterministic parity automaton for the same language, and the factor by which it can be larger is bounded by the number of colors.
This bound is also tight:
\begin{proposition}[Appears to not have been stated previously elsewhere]
Let $\inf$ be the function mapping a sequence to the set of elements occurring infinitely often in the sequence.
For every $k \in \NN$, the language $L^k = \{w \in \{1, \ldots, k\}^\omega \mid \min(\inf(w)) \text{ is even}\}$ can be represented by a deterministic parity automaton with a single state and $k$ colors, but every COCOA for the same language needs at least $k$ levels (with one state on each level). 
\end{proposition}
\begin{proof}
A deterministic parity automaton with a single state can be built with self-loops for all letters that use the letter as the respective color. The existing procedure for translating a DPW to a COCOA \cite{DBLP:conf/fsttcs/EhlersS22} then builds a COCOA $A_1, \ldots, A_k$ for this language in which on each level $i$, the words ending with $(\{i, \ldots, k\})^\omega$ are accepted. Overall, we have a blow-up by a factor of $k$, while $k$ is the number of colors in the deterministic parity automaton that we start with. 
\end{proof}

\paragraph{LTL $\rightarrow$ COCOA:}
Before discussing that COCOA can also be more concise than DPW, we look at an area in which they have the same conciseness. In particular, a translation from LTL to automata has the same worst-case blow-up lower bound for DPW and COCOA, namely doubly-exponential.

This follows from an existing proof of the doubly-exponential lower bound for translating from LTL to deterministic Büchi automata (whenever possible). Kupferman and Rosenberg gave multiple versions of such proofs for the cases of fixed and non-fixed alphabets \cite{DBLP:conf/mochart/KupfermanR10}. All proofs have in common that a family of languages is built that has a doubly-exponential number of residual languages (in the sizes of the LTL formula). This can be seen from the fact that their languages only contain words that end with $\#^\omega$ for some character $\#$ in the alphabet, and hence only a doubly-exponential blow-up in the number of residual languages can cause the automata to be so big.

\looseness-1 The complements of these languages are representable by co-Büchi automata. Furthermore, minimal \hdcw{} are \emph{semantically deterministic}, meaning that for each state $q$ in the automaton $A = (Q,\Sigma,\delta,q_0)$ reachable under a prefix word $\tilde w$, we have $\mathcal{L}((Q,\Sigma,\delta,q)) = \{w \in \Sigma^\omega \mid \tilde w w \in \mathcal{L}(A)\}$. If there is a doubly-exponential number of residual languages in $A$, we have that $A$ then needs at least a doubly-exponential number of states.
As a consequence, COCOA for these languages also need to be of doubly-exponential size, as a COCOA for a co-Büchi language consists of only a single \hdcw{} for the language.

\paragraph{COCOA conciseness over DPW:}
Let us now identify if and how COCOA can be more concise than DPWs.
For starters, it was shown that \hdcw{} can be exponentially more concise than deterministic co-Büchi automata \cite{DBLP:conf/icalp/KuperbergS15}. 
Since parity automata are co-Büchi type \cite{DBLP:journals/ijfcs/KupfermanMM06}, deterministic co-Büchi word automata cannot be less concise than deterministic parity automata.
Since furthermore COCOA for co-Büchi languages consist of a single history-deterministic co-Büchi automaton, we overall obtain that COCOA can be exponentially more concise than DPW.

We can also employ some existing results for showing that COCOA cannot be doubly-exponentially more concise than deterministic parity automata:

\begin{proposition}[Already appearing in abbreviated form in \cite{DBLP:journals/corr/abs-2410-01021} based on remarks in \cite{DBLP:conf/tacas/EhlersK24}]
\label{proposition:COCOAtoDPWExponentialIsEnough}
Let $({A}_1, \ldots, {A}_k)$ be a COCOA. There exists a deterministic parity automaton for the same language that has a number of states that is exponential in $|{A}_1|+\ldots+|{A}_k|$.
\end{proposition}
\begin{proof}
Translating a non-deterministic co-Büchi automaton to a deterministic co-Büchi automaton can be performed with an exponential blow-up \cite{DBLP:conf/icalp/BokerKR10} using the Miyano-Hayashi construction \cite{MIYANO1984321}. Doing so for each automaton in the COCOA yields a sequence of deterministic co-Büchi automata $\mathcal{D}_1, \ldots, \mathcal{D}_k$, where for each $1 \leq j \leq k$, we have $|\mathcal{D}_j| \leq 3^{|A_j|}$.

Let for each $1 \leq j \leq k$ be $\mathcal{D}_j = (Q^j,\Sigma,\delta^j,q^j_0)$. We can construct a deterministic parity automaton $\mathcal{P} = (Q^P,\Sigma,\delta^P,q^P_0)$ for the language of the COCOA as follows (using a construction from \cite{DBLP:journals/corr/abs-2410-01021}):
\allowdisplaybreaks
\begin{align*}
Q^P & = Q^1 \times \ldots \times Q^k \\
\delta^P((q^1, \ldots, q^k),x) & = ((q'^1, \ldots, q'^k),c) \text{ s.t. } \exists c^1, \ldots, c^k \in \NN. (q'^1,c^1) \in \delta^1(q^1,x), \ldots,\\
& \quad \quad (q'^k,c^k) \in \delta^k(q^k,x), c = \min(\{k \} \cup \{ j \in \{0, \ldots, k-1\} \mid c^{j+1}=2 \}) \\
q^P_0 & = (q^1_0, \ldots, q^k_0)
\end{align*}
To see that $\mathcal{P}$ has the right language, assume that for some word $w$, its natural color is $j$ for some $0 \leq j \leq k$. Then, all automata $\mathcal{D}_1, \ldots, \mathcal{D}_j$ accept the word while the automata $\mathcal{D}_{j+1}, \ldots, \mathcal{D}_k$ reject the word. Since $\mathcal{P}$ simulates all these automata in parallel, infinitely often the color $c$ along transitions in the run for $w$ will be $j$, but only finitely often the color will be in $\{0, \ldots, j-1\}$. This means that $\mathcal{P}$ accepts $w$ if and only if $j$ is even, which proves that $\mathcal{P}$ has the right language.

We have that $|\mathcal{P}| \leq |\mathcal{D}_1| \cdot \ldots \cdot |\mathcal{D}_k| \leq 3^{|{A}_1|} \cdot \ldots \cdot 3^{|{A}_k|} = 3^{|{A}_1| + \ldots + |{A}_k|} $. Overall, the blow-up of the translation is hence exponential.
\end{proof}

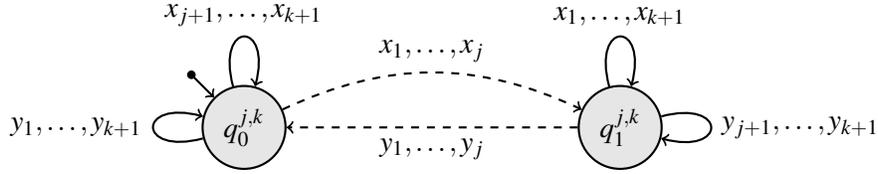
\begin{figure}
\centering\begin{tikzpicture}

\node[state,thick,fill=black!10!white] (a) at (0,0) {$q^{j,k}_0$};
\node[state,thick,fill=black!10!white] (b) at (5,0) {$q^{j,k}_1$};

\draw[thick,->] (a) to[loop above] node[above] {$x_{j+1}, \ldots, x_{k+1}$} (a);
\draw[thick,->] (a) to[loop left] node[left] {$y_{1}, \ldots, y_{k+1}$} (a);

\draw[thick,->] (b) to[loop above] node[above] {$x_{1}, \ldots, x_{k+1}$} (b);
\draw[thick,->] (b) to[loop right] node[right] {$y_{j+1}, \ldots, y_{k+1}$} (b);

\draw[thick,->,dashed] (a) to[bend left=25] node[above] {$x_{1}, \ldots, x_j$} (b);
\draw[thick,->,dashed] (b) to[bend left=0] node[below] {$y_{1}, \ldots, y_j$} (a);

\draw[fill=black] ($(a)+(-0.7,0.7)$) circle (0.05cm);
\draw[->,thick] ($(a)+(-0.7,0.7)$) -- (a);

\end{tikzpicture}
\caption{A deterministic co-Büchi automaton (parametrized for some $k \in \NN$ and $1 \leq j \leq n$) for the co-Büchi languages used in the proof of Theorem~\ref{thm:conciseNessWithSingleSuffixLanguage}}
\label{fig:firstResultSingleLevelDCW}
\end{figure}

So at a first glance, the conciseness of COCOA over DPW has been characterized to precisely singly-exponential.
What cannot be easily derived from existing results, however, is why exactly a COCOA can be exponentially more concise than a deterministic parity automaton. In particular, it may be possible that 
there are also factors other than the conciseness of history-deterministic co-Büchi automata that contribute to the  conciseness of COCOA, but they do not \emph{stack}.

It turns out that this is the case, as we show next. Even in the case that the co-Büchi languages on each level of a COCOA are representable as two-state deterministic co-Büchi automata, a parity automaton for the represented language may need exponentially more states.

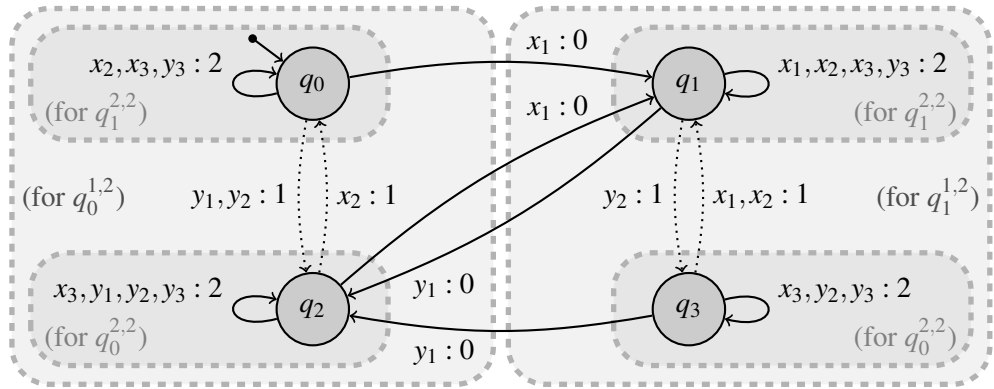
\begin{figure}
\centering\begin{tikzpicture}

\draw[line width=2pt,dashed,rounded corners=0.5cm,color=black!30!white,fill=black!5!white] (-4,-4) rectangle (2.4,1);
\draw[line width=2pt,dashed,rounded corners=0.5cm,color=black!30!white,fill=black!5!white] (2.6,-4) rectangle (9,1);

\draw[line width=2pt,dashed,rounded corners=0.5cm,color=black!30!white,fill=black!10!white] (-3.75,-3.75) rectangle +(4.75,1.5);
\draw[line width=2pt,dashed,rounded corners=0.5cm,color=black!30!white,fill=black!10!white] (-3.75,-0.75) rectangle +(4.75,1.5);
\draw[line width=2pt,dashed,rounded corners=0.5cm,color=black!30!white,fill=black!10!white] (4,-3.75) rectangle +(4.75,1.5);
\draw[line width=2pt,dashed,rounded corners=0.5cm,color=black!30!white,fill=black!10!white] (4,-0.75) rectangle +(4.75,1.5);

\node[anchor=west,color=black!70!white] at (-4,-1.5) {(for $q^{1,2}_0$)};
\node[anchor=east,color=black!70!white] at (9,-1.5) {(for $q^{1,2}_1$)};
\node[anchor=south west,color=black!50!white] at (-3.7,-3.75) {(for $q^{2,2}_0$)};
\node[anchor=south west,color=black!50!white] at (-3.7,-0.75) {(for $q^{2,2}_1$)};
\node[anchor=south east,color=black!50!white] at (8.7,-3.75) {(for $q^{2,2}_0$)};
\node[anchor=south east,color=black!50!white] at (8.7,-0.75) {(for $q^{2,2}_1$)};

\node[state,thick,fill=black!20!white] (a) at (0,0) {$q_0$};
\node[state,thick,fill=black!20!white] (b) at (5,0) {$q_1$};
\node[state,thick,fill=black!20!white] (c) at (0,-3) {$q_2$};
\node[state,thick,fill=black!20!white] (d) at (5,-3) {$q_3$};

\draw[thick,->] (a) to[loop left] node[left,yshift=0.25cm] {$x_2, x_3, y_3: 2$} (a);
\draw[thick,->] (b) to[loop right] node[right,yshift=0.25cm] {$x_1, x_2, x_3, y_3: 2$} (b);
\draw[thick,->] (c) to[loop left] node[left,yshift=0.25cm] {$x_3, y_1, y_2, y_3: 2$} (c);
\draw[thick,->] (d) to[loop right] node[right,yshift=0.25cm] {$x_3, y_2, y_3: 2$} (d); 

\draw[thick,->,pos=0.7] (a) to[bend left=10] node[above] {$x_{1}: 0$} (b);
\draw[thick,->] (c) to[bend left=10] node[above left=-1mm,pos=0.83] {$x_{1}: 0$} (b);
\draw[thick,->] (b) to[bend left=10] node[below right=-1mm,pos=0.83] {$y_{1}: 0$} (c);
\draw[thick,->,pos=0.7] (d) to[bend left=10] node[below] {$y_{1}: 0$} (c);

\draw[thick,->,dotted] (c) to[bend right=10] node[right] {$x_{2}: 1$} (a);
\draw[thick,->,dotted] (a) to[bend right=10] node[left] {$y_1, y_{2}: 1$} (c);
\draw[thick,->,dotted] (d) to[bend right=10] node[right] {$x_1, x_{2}: 1$} (b);
\draw[thick,->,dotted] (b) to[bend right=10] node[left] {$y_{2}: 1$} (d);

\draw[fill=black] ($(a)+(-0.8,0.6)$) circle (0.05cm);
\draw[->,thick] ($(a)+(-0.8,0.6)$) -- (a);

\end{tikzpicture}
\caption{A minimal DPW for $\mathcal{L}(\mathcal{C}^2)$ with a marking of how the states map to combinations of states in a COCOA for the same language}
\label{fig:dpaForC2}
\end{figure}

\begin{theorem}
\label{thm:conciseNessWithSingleSuffixLanguage}
There exists a family of COCOA $\mathcal{C}^1, \mathcal{C}^2, \ldots$ for which for each COCOA $\mathcal{C}^k = (A^k_1, \ldots, A^k_k)$, we have that for all $1 \leq j \leq k$, the history-deterministic co-Büchi automaton ${A}^k_j$ has only two states, the language of $\mathcal{C}^k$ only has a single residual language, and every deterministic parity automaton $\mathcal{P}^k$ for the language of $\mathcal{C}^k$ needs at least $2^k$ states.
\end{theorem}
For the proof of this theorem, we first define a suitable family of languages. 
\begin{definition}
For every $k \in \mathbb{N}$, we define $\mathcal{C}^k = (A^k_1, \ldots, A^k_k)$ so that the co-Büchi automata in the COCOA have the joint alphabet $\Sigma^k = \{x_1, \ldots, x_{k+1}, y_1, \ldots, y_{k+1}\}$ and such that for each $1 \leq j \leq k$, a deterministic co-Büchi automaton for $A^k_j$ can be given as in Figure~\ref{fig:firstResultSingleLevelDCW}.
\end{definition}
Intuitively, every automaton $A^k_i$ in a COCOA $\mathcal{C}^k$ accepts those words in which either the letters $x_1 \ldots x_i$ appear only finitely often or the letters $y_1 \ldots y_i$ appear only finitely often. Figure~\ref{fig:dpaForC2} depicts a minimally sized DPW $\mathcal{P}^2$ for $\mathcal{L}(\mathcal{C}^2)$ and provides some intuition on why a DPW for such a COCOA may need to be large: in order to ensure that words are accepted by the DPW that are rejected by $A^2_1$, the DPW's state set needs to be split into those states corresponding to state $q_0^{1,2}$ (on the left) and those corresponding to $q_1^{1,2}$ (on the right), so that a run switching between these infinitely often is accepting. This is implemented by the transitions between the left and right parts of the the DPW having a color of $0$, which is then the dominating color of the run. Within \emph{each} of these separate state sets, however, we also need a split between states corresponding to $q_0^{2,2}$ (the bottom two states in the DPW) and those corresponding to $q_1^{2,2}$ (the top two states in the DPW) to detect when a word should be rejected by the DPW due to it not being accepted by $A^2_2$. 
Transitions between bottom and top states have a color of $1$ to implement that the word is rejected by $\mathcal{P}^2$ if the word is rejected by $A^2_2$ (but accepted by $A^2_1$).
Such a nesting of states from different co-Büchi automata in $\mathcal{C}^k$ is indeed unavoidable, as we show next in order to prove Theorem~\ref{thm:conciseNessWithSingleSuffixLanguage}.

We employ ideas from the study of \emph{rerailing automata} \cite{DBLP:journals/corr/abs-2503-08438}, which generalize deterministic parity automata. In particular, we study how strongly connected components in a parity automaton for $\mathcal{L}(\mathcal{C}^k)$ need to be \emph{nested}. The main observation used for proving Theorem~\ref{thm:conciseNessWithSingleSuffixLanguage} that can be obtained in this way is captured in the following lemma:

\begin{lemma}
\label{lem:One}
Let $(Q',\delta')$ be a strongly connected component in $\mathcal{P}^k$ consisting only of reachable states and for some $1 \leq i < k$ and $1 \leq j < k$, we have that for any word $w$ in which only letters from $x_i, \ldots, x_{k+1}, y_j, \allowbreak{} \ldots, \allowbreak{} y_{k+1}$ occur, a run for $w$ starting in any state in $Q'$ stays in $(Q',\delta')$.

Then, we have that there are disjoint reachable SCCs $(Q'^x,\delta'^x)$ and $(Q'^y,\delta'^y)$ within $(Q',\delta')$ s.t.
\begin{itemize}
\item for any word $w'$ with only letters from $x_{\max(i,j)+1} \ldots x_{k+1}, y_j, \ldots, y_{k+1}$, any run from a state $q \in Q'^x$ for $w'$ stays in $(Q'^x,\delta'^x)$, and
\item for any word $w'$ with only letters from $x_{i} \ldots x_{k+1}, y_{\max(i,j)+1}, \ldots, y_{k+1}$, any run from a state $q \in Q'^y$ for $w'$ stays in $(Q'^y,\delta'^y)$.
\end{itemize}
\end{lemma}
\begin{proof}
We can find the SCC $(Q'^x,\delta'^x)$ as follows:
Consider the set of transitions $T^x$ in $(Q',\delta')$ for letters from $x_{\max(i,j)+1} \ldots x_{k+1}, y_j, \ldots, y_{k+1}$. We use a subset of $T^x$ that forms a transition set of an SCC  as $\delta'^x$. Such a subset has to exist as all transitions from states in $Q'$ for letters in the considered letter set stay in $Q'$, and $Q'$ together with $T^x$ decomposes into SCCs.
We find the SCC $(Q'^y,\delta'^y)$ in the same way but for the letters $x_{i} \ldots x_{k+1}, y_{\max(i,j)+1}, \ldots, \allowbreak{} y_{k+1}$.

The SCCs $(Q'^x,\delta'^x)$ and $(Q'^y,\delta'^y)$ have the needed property: all outgoing transitions for letters in the considered character sets are within $\delta'^x$/$\delta'^y$, respectively, as they consist of all transitions for the respective characters within the SCCs, and due to how they were chosen, there are no outgoing transitions in $\mathcal{P}^k$ for the respective letter set.

To see that $(Q'^x,\delta'^x)$ and $(Q'^y,\delta'^y)$ are disjoint, 
consider first a word $w^x$ containing all letters from $x_{\max(i,j)+1} \ldots x_{k+1}, y_j, \ldots, y_{k+1}$ infinitely often and for which from some $q'^x \in Q'^x$, a run for $w^x$ takes all transitions in $\delta'^x$ infinitely often. Since  $(Q'^x,\delta'^x)$ is an SCC and contains transitions for all these letters, such a word has to exist.
Note that by the definition of $\mathcal{C}^k$, we have that $w^x$ is in the language of $\mathcal{C}^k$ if and only if $\max(i,j)$ is even.
We can build a similar word $w^y$ for $x_{i} \ldots x_{k+1}, y_{\max(i,j)+1}, \ldots, y_{k+1}$. It is also in the language of $\mathcal{C}^k$ if and only if $\max(i,j)$ is even.

If $(Q'^x,\delta'^x)$ and $(Q'^y,\delta'^y)$ would overlap, we could build a word/run combination $w^\mathit{mix}$/$\pi^\mathit{mix}$ from $w^x$ and $w^y$ by 
taking the prefix run/word of $w^x$ until reaching the joint state $q_\mathit{mix} \in Q'^x \cap Q'^y$, removing the stem of $w^y$ (i.e., the characters until when the respective run reaches $q_\mathit{mix}$), and then
switching between the words whenever $q_\mathit{mix}$ is reached along the run for $w^\mathit{mix}$. The resulting word $w^\mathit{mix}$ contains all letters from $x_{i} \ldots x_{k+1}, y_j, \ldots, y_{k+1}$ infinitely often and the run for the word takes all transitions in $\delta'^x \cup \delta'^y$ infinitely often. This means that the dominating color of the run of $w^\mathit{mix}$ is the least dominating color of runs induced by $w^x$ and $w^y$, respectively.

By the definition of $\mathcal{C}^k$, whether $w^\mathit{mix}$ is in the language of $\mathcal{C}^k$ needs to differ, however, from whether $w^x$ and $w^y$ are in the language of $\mathcal{C}^k$, as $w^\mathit{mix}$ is in $\mathcal{C}^k$ if and only if $\max(i,j)$ is odd. Hence, to avoid either $w^\mathit{mix}$, $w^x$, or $w^y$ to be recognized with a color that has the wrong evenness, we have that $Q'^x$ and $Q'^y$ need to be disjoint.
\end{proof}

This lemma can be used in an induction argument over the size of $\mathcal{P}^k$:
\begin{lemma}
\label{lem:nofStates}
Let $(Q',\delta')$ be a strongly connected component in $\mathcal{P}$ such that from any state $q \in Q'$, for any word $w$ in which only letters from $x_i, \ldots, x_{k+1}, y_j, \ldots, y_{k+1}$ occur, a run for $w$ starting in $q$ stays in $(Q',\delta')$ (for some $1 \leq i \leq k$ and $1 \leq j \leq k$). %
We have that $Q'$ is of size at least $2^{k-\max(i,j)}$.
\end{lemma}
\begin{proof}
We prove the claim by induction over $\max(i,j)$, starting from the case $\max(i,j)=k$ and progressing backwards.
For the induction basis ($\max(i,j)=k$), this claim is trivially true, as at least one state is needed in $(Q',\delta')$.

For the induction step, consider a concrete combination of $(i,j)$ with $i < k$ and $j<k$ (so that the induction basis does not apply). Lemma \ref{lem:One} states that there are distinct sub-SCCs $(Q'^x,\delta'^x)$ and $(Q'^y,\delta'^y)$ within $(Q',\delta')$ for letters from $x_{\max(i,j)+1} \ldots x_{k+1}, y_j, \ldots, y_{k+1}$ and $x_{i} \ldots x_{k+1}, y_{\max(i,j)+1}, \ldots, y_{k+1}$, respectively. By the induction hypothesis, these each have sizes of $2^{k-\max(i,j)-1}$. As $(Q',\delta')$ has both of these as distinct sub-SCCs, $Q'$ needs to have at least $2^{k-\max(i,j)}$ states.
\end{proof}
We are now ready to prove Theorem~\ref{thm:conciseNessWithSingleSuffixLanguage}. Note that it has not been proven yet that $\mathcal{C}^k$ is actually a canonical COCOA of the language it represents, which requires that every word is accepted with its natural color w.r.t.~the language of $\mathcal{C}^k$. Hence, the following proof starts with establishing this fact.
\begin{proof}[Proof of Theorem~\ref{thm:conciseNessWithSingleSuffixLanguage}]
We first prove that $\mathcal{C}^k$ is the COCOA of some language (for every $k \in \NN$).
To see this, consider first some word $w$ that is rejected by $A_1^k$. Then, 
both $x_1$ and $y_1$ appear in the word infinitely often. Injecting additional letters does not change that the word is rejected, and hence words rejected by $A_1^k$ have a natural color of $0$.

For the other automata, we show by induction that if an automaton ${A}^k_i$ is the one with smallest index accepting some word $w$, then the word has a natural color of $i$ w.r.t.~the language of $\mathcal{C}^k$. 
So let us assume that $w$ is accepted by $A_i^k$ but rejected by ${A}_{i+1}^k$ (if $i<k$). Then the word either contains $x_{i+1}$ infinitely often and all characters $x_1, \ldots, x_{i}$ only finitely often, or $y_{i+1}$ infinitely often  and all characters $y_1, \ldots, y_{i}$ only finitely often.
Any injection either maintain this property (hence keeping whether the word is in the language of $\mathcal{C}^k$ or injects characters from $x_1, \ldots, x_{i}$ or $y_1, \ldots, y_{i}$ infinitely often, and then the resulting word has a natural color that is strictly smaller.

For proving the size bound, applying Lemma~\ref{lem:nofStates} on $x_1, \ldots, x_{k+1}, y_1, \ldots, y_{k+1}$ and any SCC of $(Q,\delta)$ without outgoing edges yields that at least $2^{k}$ many states are needed for $\mathcal{P}^k$. Note that such an SCC always exists.
\end{proof}

We note that for the family of languages defined in this section, the size bound of Theorem~\ref{thm:conciseNessWithSingleSuffixLanguage} is actually tight, as by generalizing the construction depicted in Figure~\ref{fig:dpaForC2}, we can obtain parity automata $\mathcal{P}^k$ of size exactly $2^{k}$.

\section{COCOA disjunction/conjunction can cause an exponential blow-up}

We have seen in the previous section that COCOA can be exponentially more concise than deterministic parity automata. 
But how \emph{brittle} is this conciseness? In particular, can it be that a language can be represented concisely with COCOA (when compared to a DPW representation), but when processing the language, conciseness is shattered by the operation performed on the language? 
In turns out that this is indeed the case when considering conjunction and disjunction operations on COCOA, as we show in this section. 
We define two families of languages that can be concisely represented and prove that when taking their conjunction or disjunction, an exponential blow-up is unavoidable for this family. In contrast, conjunction or disjunction can be performed with polynomial blow-up when using a DPW representation for this family of languages.

{
We note that the blow-up is unrelated to any automaton size increase potentially caused by disjunction or conjunction operations on \hdcw{}, of which the COCOA are composed. 
Rather, the change in conciseness is caused by a restructuring of how the language to represent is mapped to the COCOA levels. We also note that in the general case, taking the conjunction or disjunction of DPWs has an unavoidable exponential blow-up \cite{DBLP:conf/lpar/Boker18}. %
}

We start by introducing the first family of languages $\{L^k\}_{k \in \NN}$ that have 
COCOA of size polynomial in $k$, but for which the number of residual languages is exponential in $k$. 

\begin{definition}
\label{def:toyLanguage}
Let $k \in \NN$ be given. We set $\Sigma = \{X_1, \ldots, X_k, Y_1, \ldots, Y_k, a_0, \ldots, a_{4k-1}\}$ and define $L^k = \mathcal{L}(L_1^k, \allowbreak{} \ldots, \allowbreak{} L_n^k)$ for the following sequence of language $L^k_1, \ldots, L^k_k$, where $1 \leq i \leq k$:
\begin{align*} 
L^k_i & = ((\Sigma \setminus \{ X_i \}) + X_i (\Sigma \setminus \{ X_i \})^* X_i )^* (a_0 + \ldots + a_{4k-2i+1} )^\omega + \Sigma^* (a_0 + \ldots + a_{4k-2i} )^\omega
\end{align*}
\end{definition}
Each language $L_i^k$ only includes words that eventually only contain lower-case letters. Which such words are in the language only depends on which lower-case letters are infinitely often contained, and their order does not matter. If the number of $X_i$ letters at the beginning of the word is even, then the set of characters that can appear infinitely often in the word is slightly larger by also including $a_{4k-2i+1}$. For all $L^k_i$, the set of letters that may occur infinitely often is strictly larger than for $L^k_{i+1}$. Whether the number of $X_i$ letters in a word is even or odd is only relevant for $L^k_i$, but not for $L^k_j$ for $i \neq j$.
We will next show that:
\begin{itemize}
\item Each language $L^k_i$ is a co-Büchi language, and there exists a deterministic co-Büchi automaton for $L^k_i$ with 2 states.
\item $\Sigma^\omega$ has words with natural colors of $0\ldots k$ and $L_i^k$ accepts exactly the words with a natural color of $i$ or more (w.r.t.~$L^k$) -- hence, the co-Büchi automata for $L^k_1, \ldots, L^k_k$ together form a valid COCOA.
\end{itemize}

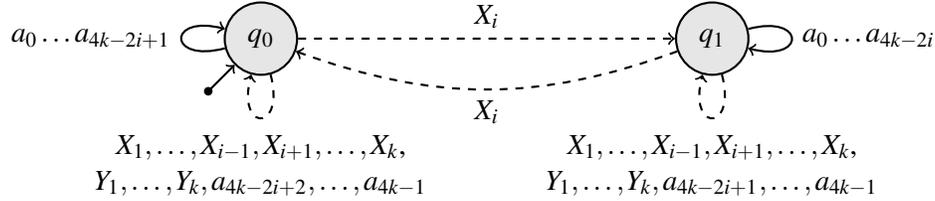
\begin{figure}
\centering
\begin{tikzpicture}

\node[state,thick,fill=black!10!white] (a) at (0,0) {$q_0$};

\node[state,thick,fill=black!10!white] (b) at (6,0) {$q_1$};

\draw[fill=black] ($(a)+(-0.7,-0.7)$) circle (0.05cm);
\draw[->,thick] ($(a)+(-0.7,-0.7)$) -- (a);

\draw[->,thick,dashed] (a) to[bend left=00] node[above] {$X_i$} (b);
\draw[->,thick,dashed] (b) to[bend left=20] node[below] {$X_i$} (a);
\draw[->,thick,dashed] (a) to[loop below] node[below] {
\begin{tabular}{c}$X_1, \ldots, X_{i-1}, X_{i+1}, \ldots, X_k$, \\ $Y_1, \ldots, Y_k, a_{4k-2i+2}, \ldots, a_{4k-1}$ \end{tabular}} (a);
\draw[->,thick] (a) to[loop left] node[left] {$a_0 \ldots a_{4k-2i+1}$} (a);

\draw[->,thick,dashed] (b) to[loop below] node[below] {
\begin{tabular}{c}$X_1, \ldots, X_{i-1}, X_{i+1}, \ldots, X_k$, \\ $Y_1, \ldots, Y_k, a_{4k-2i+1}, \ldots, a_{4k-1}$ \end{tabular}} (b);
\draw[->,thick] (b) to[loop right] node[right] {$a_0 \ldots a_{4k-2i}$} (b);
\end{tikzpicture}
\caption{A deterministic co-Büchi automaton for the language $L^k_i$. Rejecting transitions are dashed.}
\label{def:toyLanguageAutomaton}
\end{figure}

\begin{lemma}
\label{lem:gfgcoBuchiSizes}
Let $L^k_i$ be a language as defined in Def.~\ref{def:toyLanguage}. There exists a deterministic co-Büchi automaton with transition-based acceptance for $L^k_i$ with two states.
\end{lemma}
\begin{proof}
The automaton shown in Figure~\ref{def:toyLanguageAutomaton} accepts the desired language.
\end{proof}

\begin{lemma}
\label{lem:correctNaturalColors}
Let $k \in \NN$ and $L_1^k, \ldots, L_k^k$ be languages defined for $k \in \NN$ according to Def.~\ref{def:toyLanguage}. 
We have that every language $L^k_i$ contains exactly the words that have a natural color of $i$ regarding $L^k$.
\end{lemma}

Let us now define the family of languages to combine the COCOA for $L^k$ with.

\begin{definition}
\label{def:toyLanguageB}
Let $k \in \NN$ be given. We set $\Sigma = \{X_1, \ldots, X_k, Y_1, \ldots, Y_k, a_0, \ldots, a_{4k-1}\}$ 
and define $\hat L^k = \mathcal{L}(\hat L^k_1, \allowbreak{} \ldots, \allowbreak{} \hat L^k_n)$ for the following sequence of languages, where $1 \leq i \leq k$:
\begin{align*} 
\hat L^k_i & = ((\Sigma \setminus \{ Y_i \}) + Y_i (\Sigma \setminus \{ Y_i \})^* Y_i )^* (a_{2i-2} + \ldots + a_{4k-1} )^\omega + \Sigma^* (a_{2i-1} + \ldots + a_{4k-1} )^\omega
\end{align*}
\end{definition}

Note that the properties of $L^k$ established in Lemma~\ref{lem:gfgcoBuchiSizes} and Lemma~\ref{lem:correctNaturalColors} carry over to $\hat L^k$ as well, as the languages only differ by swapping the roles of the letters $\{X_i\}_{1 \leq 1 \leq k}$ and $\{Y_i\}_{1 \leq 1 \leq k}$ as well as {swapping the letters $a_i$ and  $a_{4k-i-1}$ for each $0 \leq i < 2k$}.

Let in the following $L'^k = L^k \cap \hat L^k$. 
We will next analyze how big a COCOA for $L'^k$ needs to be and in this way shed light on how big the conjunction of COCOA for $L^k$ and $\hat L^k$ need to be.
To perform this analysis, we consider the language intersections $L^k_i \cap \hat L^k_j$ (for $1 \leq i \leq k$ and $1 \leq j \leq k$) and show how a COCOA for $L'^k$ can be built from disjunctions of some co-Büchi automata for $L^k_i \cap \hat L^k_j$.

\begin{lemma}
\label{lem:movingLemma}
Let $w \in \Sigma^\omega$ be a word. There exists a unique greatest index pair $(i,j) \in \{0, \ldots, k\}^2$ such that
$w \in L^k_i \cap \hat L^k_j$, i.e., we have $w \in L^k_i \cap \hat L^k_j$ and for all $(i',j') \in \{0, \ldots, k\}^2$ such that $w \in L^k_{i'} \cap \hat L^k_{j'}$, we have that $i' \leq i$ and $j' \leq j$.

Furthermore, for every pair $(i', j')$ with $i' \leq i$ and $j' \leq j$, there exists an extension $w'$ of $w$ such that $(i', j')$ is the unique greatest index pair such that $w' \in L^k_{i'} \cap \hat L^k_{j'}$.
\end{lemma}
\begin{proof}
For the first half, first of all note that $w \in L^k_0 \cap \hat L^k_0$ by definition as both $L^k_0$ and $\hat L^k_0$ {contain all infinite words over $\Sigma$.}
Then, let $K$ be the set of elements $(i,j)$ such that we have $w \in L^k_i \cap \hat L^k_j$.
If we have $(i,j) \in K$ and $(i',j') \in K$ for some such pairs, this means that $w \in L^k_i$, $w \in \hat L^k_j$, $w \in L^k_{i'}$, and $w \in \hat L^k_{j'}$, so we then also have $(\max(i,i'),\max(j,j')) \in K$. So we cannot have that both $(i,j)$ and $(i',j')$ are incomparable (with respect to element-wise comparison) maximal elements in $K$, as otherwise $(\max(i,i'),\max(j,j'))$ is another element in $K$, contradicting the assumption that both $(i,j)$ and $(i',j')$ are incomparable maximal elements of $K$. %

For the second half of the claim, let $w$ be given, let $(i,j)$ be the (unique) maximal level in $K$, and $(i',j')$ be such that $i' \leq i$ and $j' \leq j$. By injecting infinitely often $a_{4k-2i'}$ into $w$, the resulting word is in {$L^k_{i'}$} (but not in $L^k_{i'+1}$), and by injecting infinitely often $a_{2j'-1}$, the resulting word is in $\hat L^k_{j'}$ (but not in $\hat L^k_{j'+1}$). Definitions~\ref{def:toyLanguage} and \ref{def:toyLanguageB} are such that the former letter injections do not affect where in the chain $\hat L^k_1, \ldots, \hat L^k_k$ the resulting word is located, while the latter letter injections do not affect where in the chain $L^k_1, \ldots, L^k_k$ the resulting word is located. Hence, {the extended word has $(i',j')$ as the unique greatest index pair}.
\end{proof}

\begin{figure}

\begin{tikzpicture}[yscale=0.9]
\path[use as bounding box] (-7.5,0.75) rectangle (7.5,-8.6);

\path[fill=red!10!white] (-1.5,0) ..controls +(0.75,0) and +(-0.75,0) .. (0,-0.5) ..controls +(0.75,0) and +(-0.75,0) .. (1.5,0) ..controls +(-1,1) and +(1,1) .. cycle;

\path[fill=red!10!white]   (-4.5,-2) ..controls +(0.75,0) and +(-0.75,0) .. (-3,-2.5) ..controls +(0.75,0) and +(-0.75,0) .. (-1.5,-2) ..controls +(0.75,0) and +(-0.75,0) .. (0,-2.75) ..controls +(0.75,0) and +(-0.75,0) .. (1.5,-2) ..controls +(0.75,0) and +(-0.75,0) .. (3,-2.5) ..controls +(0.75,0) and +(-0.75,0) .. (4.5,-2)
-- 
(3,-1) 
..controls +(-0.75,0) and +(0.75,0) ..
(1.5,-1.5)
..controls +(-0.75,0) and +(0.75,0) ..
(0,-2.5)
..controls +(-0.75,0) and +(0.75,0) ..
(-1.5,-1.5)
..controls +(-0.75,0) and +(0.75,0) ..
(-3.0,-1)
;

\path[fill=red!10!white] (-6,-3) 
..controls +(0.75,0) and +(-0.75,0) .. 
(-4.5,-3.5)
..controls +(0.75,0) and +(-0.75,0) ..
(-3,-4.5)
..controls +(0.75,0) and +(-0.75,0) ..
(-1.5,-3.5)
--
(1.5,-3.5)
..controls +(0.75,0) and +(-0.75,0) ..
(3.0,-4.5)
..controls +(0.75,0) and +(-0.75,0) ..
(4.5,-3.5)
..controls +(0.75,0) and +(-0.75,0) ..
(6.0,-3)
--
(7.5,-4)
..controls +(-0.75,0) and +(0.75,0) ..
(6,-4.5)
..controls +(-0.75,0) and +(0.75,0) ..
(4.5,-4)
..controls +(-0.75,0) and +(0.75,0) ..
(3.0,-4.75)
..controls +(-0.75,0) and +(0.75,0) ..
(1.5,-4.5)
--
(-1.5,-4.5)
..controls +(-0.75,0) and +(0.75,0) ..
(-3,-4.75)
..controls +(-0.75,0) and +(0.75,0) ..
(-4.5,-4) 
..controls +(-0.75,0) and +(0.75,0) ..
(-6,-4.5) 
..controls +(-0.75,0) and +(0.75,0) ..
(-7.5,-4)
-- cycle
;

\path[fill=red!10!white] (-6,-5)
..controls +(0.75,0) and +(-0.75,0) ..
(-4.5,-5.5)
--
(-1.5,-5.5)
..controls +(0.75,0) and +(-0.75,0) ..
(0,-6.5)
..controls +(0.75,0) and +(-0.75,0) ..
(1.5,-5.5)
--
(4.5,-5.5)
..controls +(0.75,0) and +(-0.75,0) ..
(6,-5.25)
--
(4.5,-6.5)
..controls +(-0.75,0) and +(0.75,0) ..
(3,-6.5)
..controls +(-0.75,0) and +(0.75,0) ..
(1.5,-6.0)
..controls +(-0.75,0) and +(0.75,0) ..
(0,-6.75)
..controls +(-0.75,0) and +(0.75,0) ..
(-1.5,-6.0)
..controls +(-0.75,0) and +(0.75,0) ..
(-3,-6.5)
..controls +(-0.75,0) and +(0.75,0) ..
(-4.5,-6)
-- cycle
;

\path[fill=red!10!white] (-3,-7)
..controls +(0.75,0) and +(-0.75,0) ..
(-1.5,-7.5)
--
(1.5,-7.5)
..controls +(0.75,0) and +(-0.75,0) ..
(3,-7.25) 
..controls +(-2,-2) and +(2,-2) ..
(-3,-7)
;

\node[inner sep=2pt] (a00) at (0,0) {$L^4_0 \cap \hat L^4_0$};

\node[inner sep=2pt] (a01) at (-1.5,-1) {$L^4_0 \cap \hat L^4_1$};
\node[inner sep=2pt] (a10) at (1.5,-1) {$L^4_1 \cap \hat L^4_0$};

\node[inner sep=2pt] (a02) at (-3,-2) {$L^4_0 \cap \hat L^4_2$};
\node[inner sep=2pt] (a11) at (0,-2) {$L^4_1 \cap \hat L^4_1$};
\node[inner sep=2pt] (a20) at (3,-2) {$L^4_2 \cap \hat L^4_0$};

\node[inner sep=2pt] (a03) at (-4.5,-3) {$L^4_0 \cap \hat L^4_3$};
\node[inner sep=2pt] (a12) at (-1.5,-3) {$L^4_1 \cap \hat L^4_2$};
\node[inner sep=2pt] (a21) at (1.5,-3) {$L^4_2 \cap \hat L^4_1$};
\node[inner sep=2pt] (a30) at (4.5,-3) {$L^4_3 \cap \hat L^4_0$};

\node[inner sep=2pt] (a04) at (-6,-4) {$L^4_0 \cap \hat L^4_4$};
\node[inner sep=2pt] (a13) at (-3,-4) {$L^4_1 \cap \hat L^4_3$};
\node[inner sep=2pt] (a22) at (0,-4) {$L^4_2 \cap \hat L^4_2$};
\node[inner sep=2pt] (a31) at (3,-4) {$L^4_3 \cap \hat L^4_1$};
\node[inner sep=2pt] (a40) at (6,-4) {$L^4_4 \cap \hat L^4_0$};

\node[inner sep=2pt] (a14) at (-4.5,-5) {$L^4_1 \cap \hat L^4_4$};
\node[inner sep=2pt] (a23) at (-1.5,-5) {$L^4_2 \cap \hat L^4_3$};
\node[inner sep=2pt] (a32) at (1.5,-5) {$L^4_3 \cap \hat L^4_2$};
\node[inner sep=2pt] (a41) at (4.5,-5) {$L^4_4 \cap \hat L^4_1$};

\node[inner sep=2pt] (a24) at (-3,-6) {$L^4_2 \cap \hat L^4_4$};
\node[inner sep=2pt] (a33) at (0,-6) {$L^4_3 \cap \hat L^4_3$};
\node[inner sep=2pt] (a42) at (3,-6) {$L^4_4 \cap \hat L^4_2$};

\node[inner sep=2pt] (a34) at (-1.5,-7) {$L^4_3 \cap \hat L^4_4$};
\node[inner sep=2pt] (a43) at (1.5,-7) {$L^4_4 \cap \hat L^4_3$};

\node[inner sep=2pt] (a44) at (0,-8) {$L^4_4 \cap \hat L^4_4$};

\draw[thick] (a00) -- (a01) -- (a02) -- (a03) -- (a04) -- (a14) -- (a24) -- (a34) -- (a44);
\draw[thick] (a00) -- (a10) -- (a20) -- (a30) -- (a40) -- (a41) -- (a42) -- (a43) -- (a44);

\draw[thick] (a10) -- (a11) -- (a12) -- (a13) -- (a14);
\draw[thick] (a20) -- (a21) -- (a22) -- (a23) -- (a24);
\draw[thick] (a30) -- (a31) -- (a32) -- (a33) -- (a34);
\draw[thick] (a01) -- (a11) -- (a21) -- (a31) -- (a41);
\draw[thick] (a02) -- (a12) -- (a22) -- (a32) -- (a42);
\draw[thick] (a03) -- (a13) -- (a23) -- (a33) -- (a43);

\draw[thick,color=red] (-1.5,0) ..controls +(0.75,0) and +(-0.75,0) .. (0,-0.5) ..controls +(0.75,0) and +(-0.75,0) .. (1.5,0) node[above left,yshift=4pt,xshift=12pt,fill=red!10!white,shape=rectangle,rounded corners=0.2cm,inner sep=3pt] {$\mathcal{L}(C^4_0)$};

\draw[thick,color=red] (-3.0,-1) ..controls +(0.75,0) and +(-0.75,0) .. (-1.5,-1.5) ..controls +(0.75,0) and +(-0.75,0) .. (0,-2.5) ..controls +(0.75,0) and +(-0.75,0) .. (1.5,-1.5) ..controls +(0.75,0) and +(-0.75,0) .. (3,-1) 
node[above left,shape=rectangle,rounded corners=0.2cm,inner sep=3pt,xshift=12pt] {$\mathcal{L}(C^4_1)$};

\draw[thick,color=red] (-4.5,-2) ..controls +(0.75,0) and +(-0.75,0) .. (-3,-2.5) ..controls +(0.75,0) and +(-0.75,0) .. (-1.5,-2) ..controls +(0.75,0) and +(-0.75,0) .. (0,-2.75) ..controls +(0.75,0) and +(-0.75,0) .. (1.5,-2) ..controls +(0.75,0) and +(-0.75,0) .. (3,-2.5) ..controls +(0.75,0) and +(-0.75,0) .. (4.5,-2) node[above left,yshift=4pt,fill=red!10!white,shape=rectangle,rounded corners=0.2cm,inner sep=3pt,xshift=12pt] {$\mathcal{L}(C^4_2)$};

\draw[thick,color=red] (-6,-3) 
..controls +(0.75,0) and +(-0.75,0) .. 
(-4.5,-3.5)
..controls +(0.75,0) and +(-0.75,0) ..
(-3,-4.5)
..controls +(0.75,0) and +(-0.75,0) ..
(-1.5,-3.5)
--
(1.5,-3.5)
..controls +(0.75,0) and +(-0.75,0) ..
(3.0,-4.5)
..controls +(0.75,0) and +(-0.75,0) ..
(4.5,-3.5)
..controls +(0.75,0) and +(-0.75,0) ..
(6.0,-3)
node[above left,shape=rectangle,rounded corners=0.2cm,inner sep=3pt,xshift=12pt] {$\mathcal{L}(C^4_3)$}
;

\draw[thick,color=red] (-7.5,-4)
..controls +(0.75,0) and +(-0.75,0) ..
(-6,-4.5) 
..controls +(0.75,0) and +(-0.75,0) ..
(-4.5,-4) 
..controls +(0.75,0) and +(-0.75,0) ..
(-3,-4.75)
..controls +(0.75,0) and +(-0.75,0) ..
(-1.5,-4.5)
--
(1.5,-4.5)
..controls +(0.75,0) and +(-0.75,0) ..
(3.0,-4.75)
..controls +(0.75,0) and +(-0.75,0) ..
(4.5,-4)
..controls +(0.75,0) and +(-0.75,0) ..
(6,-4.5)
..controls +(0.75,0) and +(-0.75,0) ..
(7.5,-4)
node[above left,yshift=4pt,fill=red!10!white,shape=rectangle,rounded corners=0.2cm,inner sep=3pt,xshift=12pt] {$\mathcal{L}(C^4_4)$}
;

\draw[thick,color=red] (-6,-5)
..controls +(0.75,0) and +(-0.75,0) ..
(-4.5,-5.5)
--
(-1.5,-5.5)
..controls +(0.75,0) and +(-0.75,0) ..
(0,-6.5)
..controls +(0.75,0) and +(-0.75,0) ..
(1.5,-5.5)
--
(4.5,-5.5)
..controls +(0.75,0) and +(-0.75,0) ..
(6,-5.25)
node[above left,shape=rectangle,rounded corners=0.2cm,inner sep=3pt,xshift=20pt] {$\mathcal{L}(C^4_5)$}
;

\draw[thick,color=red] (-4.5,-6)
..controls +(0.75,0) and +(-0.75,0) ..
(-3,-6.5)
..controls +(0.75,0) and +(-0.75,0) ..
(-1.5,-6.0)
..controls +(0.75,0) and +(-0.75,0) ..
(0,-6.75)
..controls +(0.75,0) and +(-0.75,0) ..
(1.5,-6.0)
..controls +(0.75,0) and +(-0.75,0) ..
(3,-6.5)
..controls +(0.75,0) and +(-0.75,0) ..
(4.5,-6.5)
node[above left,yshift=4pt,fill=red!10!white,shape=rectangle,rounded corners=0.2cm,inner sep=3pt,xshift=20pt] {$\mathcal{L}(C^4_6)$}
;

\draw[thick,color=red] (-3,-7)
..controls +(0.75,0) and +(-0.75,0) ..
(-1.5,-7.5)
--
(1.5,-7.5)
..controls +(0.75,0) and +(-0.75,0) ..
(3,-7.25)
node[above left,shape=rectangle,rounded corners=0.2cm,inner sep=3pt,xshift=20pt] {$\mathcal{L}(C^4_7)$}

node[below left,xshift=-5pt,yshift=-5pt,fill=red!10!white,shape=rectangle,rounded corners=0.2cm,inner sep=3pt,xshift=12pt] {$\mathcal{L}(C^4_8)$}
;

\end{tikzpicture}
\caption{Overview of how the sets $\{ L^k_i \cap \hat L^k_j\}_{0 \leq i \leq k, 0 \leq j \leq k}$ (for $k=4$) compose $C^k_0, \ldots, C_{2k}^k$ in Theorem~\ref{thm:conjunctionBecomesBig}} 
\label{fig:levels}
\end{figure}
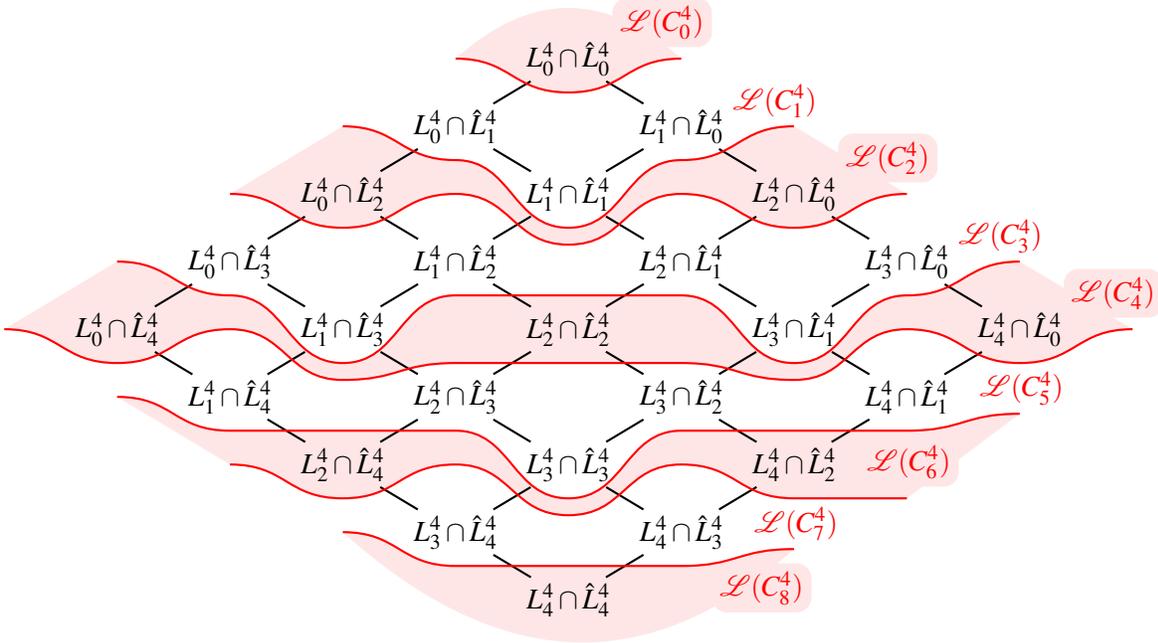

\begin{theorem}
\label{thm:conjunctionBecomesBig}
A COCOA for $L'^k = L^k \cap \hat L^k$ can be given as $\mathcal{C}^{k} = (C^k_1, \ldots, C^k_{2k})$ where for each $u \in \{0, \allowbreak{} \ldots, \allowbreak{}2k\}$, the language of $C^k_u$ is 
\begin{equation*}
\mathcal{L}(C^k_{u}) = \bigcup_{(i,j) \in \Gamma_u} L^k_i \cap \hat L^k_j
\end{equation*}
for
\begin{equation*}
\Gamma^k_u = \begin{cases}
\{(i,j) \in \{0, \ldots, k\}^2 \mid i+j=u, i \text{ is even},j \text{ is even} \} & \text{if } u \in \{0, 2, \ldots, 2k\} \\
\{(i,j) \in \{0, \ldots, k\}^2 \mid u \leq i+j \leq u+1, i \text{ is odd} \text{ or } j \text{ is odd} \} & \text{if } u \in \{1, 3, \ldots, 2k-1\}.
\end{cases}
\end{equation*}
\end{theorem}
\looseness-1 Figure~\ref{fig:levels} shows how the languages $\{ L^k_i \cap \hat L^k_j\}_{0 \leq i \leq k, 0 \leq j \leq k}$ are grouped (by performing language disjunctions) to form a COCOA for $L'^k$. Languages $L^k_i \cap \hat L^k_j$ in which both $i$ and $j$ are even form the accepting levels of the COCOA, and the languages in between are grouped into rejecting levels of the COCOA for $L'^k$.

Theorem~\ref{thm:conjunctionBecomesBig} provides a blueprint for building $\mathcal{C}^k$ from the COCOA for $L^k$ and $\hat L^k$. In particular, we can obtain $\mathcal{C}^k$ by a sequence of disjunction and conjunction operations. This characterization allows us to deduce that $\mathcal{C}^k$ must be of size exponential in $k$, as proposition~\ref{prop:exponentialSize} below shows. For it, we employ conjunction/disjunction operations for deterministic co-Büchi automata, adapted to the case of transition-based acceptance:
\begin{lemma}
\label{lem:conjunctionDisjunctionOfDCW}
Let $\mathcal{A}_1, \ldots, \mathcal{A}_n$ be deterministic co-Büchi automata over the same alphabet. We can construct a deterministic co-Büchi automaton $\mathcal{A}^\wedge$ for the conjunction of these languages of size $|\mathcal{A}_1| \cdot \ldots \cdot |\mathcal{A}_n|$, and a deterministic co-Büchi automaton $\mathcal{A}^\vee$ for the disjunction of theses languages of size $|\mathcal{A}_1| \cdot \ldots \cdot |\mathcal{A}_n| \cdot n$.
\end{lemma}

\begin{proposition}
\label{prop:exponentialSize}
Let $\mathcal{C}^k = (C^k_1, \ldots, C^k_{2k})$ be a COCOA for $L^k \cap \hat L^k$. 

A minimal deterministic co-Büchi automaton for $\mathcal{L}(C^k_i)$ for some $1 \leq i \leq 2k$ has at most $2^{2k} \cdot k$ many states.
For even $k$, we have that the \hdcw{} $C^k_k$ has at least $2^{k}$ many states. For odd $k$, we have that $C^k_{k-1}$ has at least $2^{k-2}$ many states.
\end{proposition}
\begin{proof}
Let for all languages $L^k_i$ and $\hat L^k_j$ be the respective two-state deterministic automata be denoted by $A^k_i$ and $\hat A^k_j$, which by Lemma~\ref{lem:gfgcoBuchiSizes} have two states each.

\looseness-1 For the first part, note that for all $0 \leq i \leq 2k$, the set $\Gamma^k_u$ has at most $k$ many \emph{non-dominated elements}, i.e., pairs $(i,j)$ that do not have another different pair $(i',j')$ in the set such that $i'\geq i$ and $j' \geq j$. When building $C^k_u = \bigcup_{(i,j) \in \Gamma^k_u} A^k_i \cap \hat A^k_j$, only the non-dominated pairs have to be considered, as all words accepted by $A^k_{i'} \cap \hat A^k_{j'}$ for a dominated pair $(i',j')$ are also accepted by $A^k_{i} \cap \hat A^k_{j}$ for some non-dominated pair $(i,j)$.
A deterministic co-Büchi automaton $A^k_i \cap \hat A^k_j$ for some pair $(i,j)$ only needs 4 states by Lemma~\ref{lem:conjunctionDisjunctionOfDCW}. Taking the union of $k$ many such automata yields an automaton with at most $2^{2k} \cdot k$ many states by the same lemma.

For the second part, consider the case of
$C^k_k = \bigcup_{(i,j) \in \Gamma^k_k} A^k_i \cap \hat A^k_j$ for even $k$. Here, we take the disjunction of $\frac{k}{2}$ many 4 state automata, yielding an automaton with at most $2^k$ many states. This is at the same time also the lower bound, because the number of residual languages of $\mathcal{L}(C^k_k)$ is $2^k$. This is because every language $L^k_i \cap \hat L^k_j$ for $(i,j) \in \Gamma^k_u$ has the word suffix $\tilde w = (a_{4k-2i+1} a_{2j-2})^\omega$ that is not accepted by any $L^k_{i'} \cap \hat L^k_{j'}$ with $(i',j') \in \Gamma^k_u$ as well for $(i,j) \neq (i',j')$ and that is only in $L^k_i \cap \hat L^k_j$ for prefixes for which the letters $X_i$ and $Y_j$ each occur an even number of times in the prefix. This implies that a \hdcw{} for $C^k_k$ has different residual languages for words that differ in their numbers of $X_i$ or $Y_j$ letters for $(i,j) \in \Gamma^k_k$. The number of residual languages is hence $2^k$.
As minimal canonical \hdcw{} are \emph{semantically deterministic} \cite{DBLP:journals/lmcs/RadiK22}, having $2^k$ many residual languages implies a lower bound of $2^k$ for the size of $C^k_k$.
The case for $k$ being odd is analogous.
\end{proof}
Let us finally discuss that a similar blow-up does not occur when representing $L^k$ and $\hat L^k$ as deterministic parity automata.
\begin{proposition}
\label{proposition:paritySizes}
Each of the languages $L^k$ (from Def.~\ref{def:toyLanguage}) and $\hat L^k$ (from Def.~\ref{def:toyLanguageB}) can be represented as deterministic parity automata with $2^k$ states (and not less).

There exists a deterministic parity automaton for $L^k \cap \hat L^k$ with no more than $2^{4k^2} \cdot k^{2k}$ many states.
\end{proposition}
\begin{proof}
We can build a DPW $\mathcal{P}^k = (Q,\Sigma,\delta,q_0)$ for $L^k$ with $Q = \mathbb{B}^k$, $q_0 = (0, \ldots,0)$, and for all $(b_1,\ldots, \allowbreak{} b_k)\in Q$ and $x \in \Sigma$, we have $\delta((b_1,\ldots, b_k),x) = ((b'_1, \ldots, b'_k),c)$ for
$b'_i = \neg b_i$ if $x=X_i$ and $b'_i = b_i$ (for all $1 \leq i \leq k$) otherwise. The value of $c$ in this transition is defined as:
\begin{equation*}
c = \begin{cases} 0 &\text{ if } x \in \{X_1, \ldots, X_k, Y_1, \ldots, Y_k\} \\
i &\text{ if } x=a_{4k-2i} \text{ for some } 1 \leq i \leq k \\
i &\text{ if } x=a_{4k-2i+1} \text{ and } b_i=0 \text{ for some } 1 \leq i \leq k \\
i-1 &\text{ if } x=a_{4k-2i+1} \text{ and } b_i=1 \text{ for some } 1 \leq i \leq k \\
k & \text{ otherwise}. 
\end{cases}
\end{equation*}
This DPW accepts every word with its natural language (w.r.t.~$L^k$), which we can see by the definition of $c$ mapping the transition for a character $a_j$ for some $j \in \NN$ to the index $i$ of the last automaton in the chain (or $0$ if there is no such automaton) containing words that contain this letter infinitely often. For a part of the letters, whether the respective $X_i$ symbol has been seen an even or odd number of times so far is also taken into account. For letters from $X_1, \ldots, X_k, Y_1, \ldots, Y_k$, the respective transition color is always $0$, as whenever they occur infinitely often along a word, the word is in $L^k$.

Note that $\mathcal{P}^k$ is the smallest deterministic parity automaton for $L^k$ as it has $2^k$ many states and the number of residual languages of $L^k$ is $2^k$, so it cannot be smaller. %

A similar DPW can be built from $\hat L^k$ by replacing $X_i$ characters with $Y_i$ and renumbering the indices for the $a_j$ letters.

For the DPW for $L^k \cap \hat L^k$, we employ Proposition~\ref{prop:exponentialSize} to obtain deterministic co-Büchi automata of size at most $2^{2k} \cdot k$ for each of the levels of a COCOA for $L^k$ and then build a product parity automaton of the deterministic automata as in 
Proposition~\ref{proposition:COCOAtoDPWExponentialIsEnough},
which yields a deterministic parity automaton for $L'^k$ of size at most $(2^{2k} \cdot k)^{2k} = 2^{4k^2} \cdot k^{2k}$.
\end{proof}

Proposition~\ref{proposition:paritySizes}, Proposition~\ref{prop:exponentialSize}, and Lemma~\ref{lem:gfgcoBuchiSizes} together show that while $L^k$ and $\hat L^k$ can be represented with a COCOA that is exponentially more concise than any deterministic parity automaton for these languages, exponential conciseness is lost when computing a COCOA for $L^k \cap \hat L^k$.

\begin{remark}
\label{remark:deMorgan}
Exponential conciseness can also be lost when taking the disjunction of two COCOA (instead of taking their conjunction).
\end{remark}
\begin{proof}
The languages $L^k$ and $\hat L^k$ have been defined such that COCOA representations for their complements can be obtained by adding a \hdcw{} accepting the universal language as new first automaton in the chains, moving all chain elements one element back. After taking the conjunction of the resulting COCOA for the complement language, we obtain a result COCOA in which the first automaton accepts the universal language (by the construction in Theorem~\ref{thm:conjunctionBecomesBig}). Removing it yields a COCOA for $L^k \cup \hat L^k$. Applying Proposition~\ref{prop:exponentialSize} for the conjunction automaton yields the exponential lower size bound on the COCOA for $L^k \cup \hat L^k$. For the parity automata that we compare with, complementation can be performed without blow-up by adding $1$ to each transition color.
\end{proof}

\section{Conclusion}
\label{sec:conclusion}

In this paper, we took a close look at the conciseness of chains of history-deterministic co-Büchi automata with transition-based acceptance (COCOA) over deterministic parity automata by aggregating previous results on them, deriving corollaries from them, and augmenting them with two more results that were not corollaries of existing work. 

In particular, we showed that by splitting a language to be represented into levels, COCOA can be exponentially more concise even when the automata on the individual levels are not. Secondly, we showed that exponential conciseness can be broken by performing language disjunction or conjunction. While the first of these results even holds for a language that only has a single residual language, the loss of conciseness in the second result was caused by co-Büchi automata in the resulting COCOA needing to represent an exponential number of residual languages. None of the two more complex results depend on the conciseness of history-deterministic co-Büchi automata over deterministic co-Büchi automata.

Apart from providing some insight into the capabilities and limits of COCOA as a representation for $\omega$-regular languages, our results inform future work on algorithms for performing operations on COCOA as well as future work on using COCOA for practical applications. In particular, the lower bound on conjunction/disjunction provides a baseline that future COCOA conjunction/disjunction algorithms can be compared against.

\bibliographystyle{eptcs}
\bibliography{bib}
\end{document}